\documentstyle[12pt]{article}
\normalbaselines
\begin{document}
\bibliographystyle{unsrt}

\bigskip\bigskip

INFN-NA-IV-93/30~~~~~~~~~~~~~~~~~~~~~~~~~~~~~~~~~~~~~~~~~~~~~~DSF-T-93/30

\vspace{4cm}

\begin{center}
{\LARGE \bf SOME RULES FOR POLYDIMENSIONAL SQUEEZING}\\[7mm]
V.I. Man'ko\\
{\it Dipartimento di Scienze Fisiche
Universita di Napoli "Federico II"and I.N.F.N.,Sez.di Napoli
 and \\{\it Lebedev Physics Institute, Leninsky pr.,53, Moscow,
117333 Russia}\\[5mm]
\end{center}

\newpage

\begin{abstract}

The review of the following results of the Refs.
\cite{Sem} - \cite{Ans} is presented:

For mixed state light of $N$-mode
electromagnetic field described by Wigner function which has
generic Gaussian form the photon distribution function is
obtained and expressed expliciltly in terms of Hermite
polynomials of $2N$-variables.The momenta of this distribution
are calculated and expressed as functions of matrix invariants
of the dispersion matrix.The role of new uncertainty relation
depending on photon state mixing parameter is elucidated.New
sum rules for Hermite polynomials of several variables are
found.The photon statistics of polymode even and odd
coherent light and squeezed polymode Schr\"odinger cat light
is qiven explicitly.Photon distribution for polymode squeezed
number states expressed in terms of multivariable Hermite
polynomials is discussed.
\end{abstract}

\newpage

\section{Introduction}
In the Ref.\cite{Sem} it was shown that the matrix elements of
density matrix in number state basis for polymode oscillator
are expressed in terms of Hermite polynomials of several
variables for the density operator in the canonically transformed
thermal state of the oscillator.In the recent works \cite{Ola1},
\cite{Ola2} the photon distribution function for the generic
Gaussian light described by the Wigner function which is the
most generic Gaussian in quadrature phase space was found and
expressed in terms of Hermite polynomials of 2 variables for
one-mode case \cite{Ola1} and $2N$ variables for polymode
case \cite{Ola2}.The physical meaning of mixed Gaussian state
of the light may be understood if one takes into account that
the pure multimode Gaussian state
corresponds to the generalised correlated state introduced
by Sudarshan \cite{Sud} who related those states
to the symplectic dynamical group.The mixed Gaussian states
studied above may be considered as the
mixture of generalised correlated states plus thermal noise
acting on each mode which has its own temperature.
The photon distribution function for even
and odd coherent states \cite{Dod72} or Schr\"odinger cat
states \cite{Har} subject to squeezing both in one-mode and
polymode cases has been found in Ref.\cite{Nik}.The polymode
Schr\"odinger cat states and photon distributions for light
in these states were introduced in \cite{Ans}.The aim of
this work is to give a review of the photon distribution
functions and related sum rules for one-mode and polymode
Gaussian light and for the even and odd coherent states
light using the results of \cite{Ola1} - \cite{Ans}.
It should be emphasized that the squeezed
light is worth to be used in interferometric gravitational
antennas \cite{Cav} and the even and odd coherent light
may play alternative role in gravitational wave experiment
\cite{Sol1},\cite{Sol2}.

\section{Photon distribution for polymode Gaussian light}
The mixed squeezed state of the $N$-mode light with a
{\em Gaussian\/} density
operator $\hat{\varrho}$ is described by the Wigner function
(see,for example,\cite{Dod89})
\begin{equation}
W({\bf p},{\bf q})
=(\det {\bf M})^{-\frac 12}\exp\left
[-\frac 12({\bf Q}-<{\bf Q}>){\bf M}^{-1}({\bf Q}-<{\bf Q}>)\right
],\label{1}
\end{equation}
\noindent where $2N$-dimensional vector ${\bf Q}=({\bf p},{\bf q}
)$ consists of $N$
components $p_1,...,p_N$ and $N$ components $q_1,...,q_N$.
\noindent$2N$ parameters $<p_i>$ and $<q_i>$, $i=1,2,\ldots ,N$,
combined
into vector $<{\bf Q}{\bf >}$, are the average values of the
quadratures.A real symmetric quadrature dispersion matrix
${\bf M}$ has $2N^2+N$ parameters
\begin{equation}
{\cal M}_{\alpha\beta}=\frac 12\left\langle\hat Q_{\alpha}\hat
Q_{\beta}+\hat Q_{\beta}\hat Q_{\alpha}\right\rangle -\left\langle
\hat Q_{\alpha}\right\rangle\left\langle\hat Q_{\beta}\right\rangle
,\qquad\alpha ,\beta =1,2,\ldots ,2N.\label{2}
\end{equation}
\noindent They obey certain constraints, which are nothing but the
generalized uncertainty relations \cite{Dod89}.
The photon distribution function in this state has the form
\cite{Ola2},\cite{Sem}
\begin{equation}
{\cal P}_{{\bf n}}={\cal P}_0\frac {H_{{\bf n}{\bf n}}^{
\{{\bf R}\}}({\bf y})}{{\bf n}!},\label{3}
\end{equation}
 where vector
${\bf n}$ consists of $N$ nonnegative
integers:  ${\bf n}=(n_1,n_2,\ldots ,n_N)${\bf .}
The function $H_{{\bf n}{\bf n}}^{\{{\bf R}\}}({\bf y})$ is the
Hermite polynomial of $2N$ variables.
We introduced also notations
\begin{equation}
{\bf n}!
=n_1!n_2!...n_N!.\label {4}
\end{equation}
\noindent The symmetric $2N$-dimensional matrix ${\bf R}$ and the
$2N$-dimensional
vector ${\bf y}$ are given by the relations
\begin{equation}
{\bf R}
={\bf U}^{\dagger}\left({\bf I}_{2N}-2{\bf M}\right
)\left({\bf I}_{2N}+2{\bf M}\right)^{-1}{\bf U}^{*},
\label{5}
\end{equation}

\begin{equation}
{\bf y}
=2{\bf U}^t({\bf I}_{2N}-2{\bf M})^{-1}<{\bf Q}
>,\label{6}
\end{equation}
where
the $2N$-dimensional unitary matrix

\[{\bf U}=\frac 1{\sqrt {2}}\left(\begin{array}{cc}
-i{\bf I}_N&i{\bf I}_N\\
{\bf I}_N&{\bf I}_N\end{array}
\right)\]
is introduced.The matrices ${\bf I}_{N}$ and ${\bf I}_{2N}$
are identity matrices of corresponding dimensions.

The probability to have no photons has the form
\begin{equation}
{\cal P}_0=
\left[\det\left({\bf M}+\frac 12{\bf I}_{2N}\right)\right
]^{-\frac 12}\exp\left[-<{\bf Q}>\left(2{\bf M}
+{\bf I}_{2N}\right)^{-1}<{\bf Q}>\right].\label{7}
\end{equation}
It may be shown \cite{Ola2},\cite{Nik}
that the multivariable Hermite polynomial is the even
function if the sum of its indices is the even number and
the polynomial with the odd sum of indices is the odd function.
Due to this pairity property of the polydimensional Hermite
polynomials the "diagonal"
multivariable Hermite polynomial is the even function since
the sum of its indices is always even number.Consequently the
above photon distribution function is the even function.
Thus the photon distribution function for generic mixed
Gaussian light found in\cite{Ola2} is expressed in terms of
multivariable Hermite polynomials and it depends on the
quadrature means and dispersions.The photon number means and
dipersion matrix corresponding to the found distribution (3)
are of the form
\begin{eqnarray*}
<n_j>=\frac 12(\sigma_{p_jp_j}+\sigma_{q_jq_j}-1)+
\frac 12(<p_j>^2+<q_j^2>),\\
\sigma_{n_jn_j}=\frac 12(T_j^2-2d_j-\frac 12)+
<{\bf Q}_j>{\cal M}_j<{\bf Q}_j>,
\end{eqnarray*}
where $T_j$ and $d_j$ are the trace and the determinant of
the $2$x$2$-matrix ${\cal M}_j$,describing only j-th mode,
and the 2-vector ${\bf Q}_j$ has the components $(p_j,q_j)$.

\section{Pure Polymode States}
The photon distribution for polymode squeezed correlated
state may be expressed in terms of symplectic transform
parameters relating boson operators as follows
\begin{equation}
\left(\begin{array}{c}
\hat {{\bf b}}\\
\hat {{\bf b}}^{\dagger}\end{array}
\right)=\Omega\left(\begin{array}{c}
\hat {{\bf a}}\\
\hat {{\bf a}}^{\dagger}\end{array}
\right)+\left(\begin{array}{c}
{\bf d}\\
{\bf d}^{*}\end{array}
\right),\qquad\Omega =\left(\begin{array}{cc}
\zeta&\eta\\
\eta^{*}&\zeta^{*}\end{array}
\right),\label{8}
\end{equation}
where $\Omega$ is a symplectic $2N$x$2N$-matrix  consisting of
four $N$-dimensional complex square blocks, and {\bf d} is a
complex $N$-vector.Then we have for photon distribution in
squeezed correlated polymode state $|\beta >$ labeled by
the complex number vector with $N$-components
\begin{equation}
{\cal P}_{{\bf n}}=\frac {{\cal P}_0(\beta )}{{\bf n}
!}\left|H_{{\bf n}}^{\{\zeta^{-1}\eta \}}
\left(\eta^{-1}[\beta -{\bf d}
]\right)\right|^2,\qquad {\cal P}_0(\beta )=
|{\cal F}_0(\beta )|^2,\label{9}
\end{equation}
where
\begin{equation}
{\cal F}_0(\beta )=(\det\zeta )^{-\frac 12}\exp\left
[\frac 12\beta \eta ^{*}\zeta ^{-1}\beta +
\beta ({\bf d}^{*}-\eta ^{*}\zeta ^{-1}{\bf d})+
\frac 12{\bf d}\eta ^{*}\zeta ^{-1}{\bf d}-|{\bf d}
|^2\right].\label{10}
\end{equation}
The photon doistribution function of the {\em squeezed
number state $|{\bf m}>$\/} is descibed by the formula
\begin{equation}
{\cal P}_{{\bf n}}=
|\det\zeta |^{-1}\exp\left[\mbox{Re}({\bf d}
\eta ^{*}\zeta ^{-1}{\bf d})-
2|{\bf d}|^2\right]\frac {\left|H_{{\bf n}
{\bf m}}^{\{{\bf R}\}}({\bf L})
\right|^2}{{\bf n}!{\bf m}!}.\label{11}
\end{equation}
Here {\bf m} is the label of the state, whereas {\bf n}
is a discrete vector variable.
$2N$x$2N$-matrix ${\bf R}$
and $2N$-vector {\bf L} are expressed now in
terms of blocks of matrix $\Omega$ and vector {\bf d}
as follows,
\begin{equation}
{\bf R}=\left(\begin{array}{cc}
\zeta^{-1}\eta&-\zeta^{-1}\\
-\zeta^{-1}&-\eta^{*}\zeta^{-1}\end{array}
\right),\qquad {\bf L}={\bf R}^{*}\left(\begin{array}{c}
-\zeta^{-1}{\bf d}\\
{\bf d}^{*}-\eta^{*}\zeta^{-1}{\bf d}\end{array}
\right).\label{12}
\end{equation}

\section{Even and odd coherent states}
The one-mode even and odd coherent states have been introduced
in Ref.\cite{Dod72}.The polymode even and odd coherent states
have been introduced in Ref.\cite{Ans}.The squeezed and correlated
even and odd coherent states have been introduced and studied in
Ref.\cite{Nik}.We will discuss the photon statistics of the light
in these states which are also called Schr\"odinger cat states
\cite{Har}.The multimode Schr\"odinger cat states are defined
by the relation \cite{Ans}
\begin{equation}
\mid {\bf A_{\pm}}>=N_{\pm} (\mid {\bf A}> \pm \mid -{\bf A}>),
\end{equation}
where the multimode coherent state $\mid {\bf A}>$ is
\begin{equation}
\mid {\bf A}>
=\mid \alpha_{1},\alpha_{2},\alpha_{3},......,\alpha_{n}>
=D({\bf A}) \mid {\bf 0}>,
\end{equation}
and $D({\bf A})$ is the multimode displacement operator creating
coherent state from the vacuum.The normalization constants are
\begin{eqnarray}
N_{+}&=&\frac{e^{\frac{\mid {\bf A} \mid^{2}}{2}}}
{2 \sqrt{cosh\mid {\bf A}\mid^{2}}},\nonumber\\
N_{-}&=&\frac{e^{\frac{\mid {\bf A} \mid^{2}}{2}}}
{2 \sqrt{sinh\mid {\bf A}\mid^{2}}},
\end{eqnarray}
where complex number ${\bf A}$ has the form
\begin{equation}
\mid {\bf A} \mid^{2}=\mid \alpha_{1} \mid^{2}
+\mid \alpha_{2} \mid^{2}
+........+\mid \alpha_{n} \mid^{2}
=\sum_{m=1}^{n}\mid \alpha_{m} \mid^{2}.
\end{equation}
The photon distribution function has the form \cite{Ans}
\begin{eqnarray}
P_{+} (n)&=&\frac{\mid \alpha_{1} \mid^{2n_{1}}
\mid \alpha_{2} \mid^{2n_{2}}...\mid \alpha_{n} \mid^{2n_{n}} }
{(n_{1}!)(n_{2}!)....(n_{n}!)
cosh\mid {\bf A}\mid^{2}},
{}~~~{\scriptsize n_{1}+n_{2}+....+n_{n}=2k},\nonumber\\
 P_{-} (n)&=&\frac{\mid \alpha_{1} \mid^{2n_{1}}
\mid \alpha_{2} \mid^{2n_{2}}...\mid \alpha_{n} \mid^{2n_{n}} }
{(n_{1}!)(n_{2}!)....(n_{n}!)
sinh\mid {\bf A}\mid^{2}},
{}~~~{\scriptsize n_{1}+n_{2}+....+n_{n}=2k+1},\nonumber\\
\end{eqnarray}
and the photon means corresponding to these distributions are
\begin{eqnarray}
<{\bf A_{+}}\mid n_{i} \mid {\bf A_{+}}>&=&\mid \alpha_{i}\mid^{2}
tanh\mid {\bf A}\mid^{2},\nonumber\\
<{\bf A_{-}}\mid n_{i} \mid {\bf A_{-}}>&=&\mid \alpha_{i}\mid^{2}
coth\mid {\bf A}\mid^{2}.
\end{eqnarray}
The photon number dispersion matrix has the matrix elements
\begin{eqnarray}
\sigma_{ik}^{+}&=&\mid \alpha_{i} \mid^{2}\mid \alpha_{k} \mid^{2}
sech^{2}\mid {\bf A} \mid^{2}+\mid \alpha_{i} \mid^{2}tanh
\mid {\bf A} \mid^{2}\delta_{ik},\nonumber\\
\sigma_{ik}^{-}&=&-\mid \alpha_{i} \mid^{2}\mid \alpha_{k} \mid^{2}
cosech^{2}\mid {\bf A} \mid^{2}+\mid \alpha_{i} \mid^{2}coth
\mid {\bf A} \mid^{2}\delta_{ik}.\nonumber\\
\end{eqnarray}

\section{Squeezed Schr\"odinger cat states}
Let us find out the photon statistics of squeezed polymode
Schr\"odinger cat state labeled by the complex $N$ -vector
$\beta $.To do that let us difine transition amplitude
from the polymode squeezed and correlated state $|\beta >$
to the polymode photon number state $|n>$
\begin{equation}
T_{n}(\beta )={\cal F}_0(\beta )\frac {1}{\sqrt {n!}}
H_{{\bf n}}^{\{\zeta^{-1}\eta \}}\left(\eta^{-1}[\beta -
{\bf d}]\right),
\end{equation}
where
\begin{equation}
{\cal F}_0(\beta )=(\det\zeta )^{-\frac 12}\exp\left
[\frac 12\beta\eta^{*}\zeta^{-1}\beta +
\beta ({\bf d}^{*}-\eta^{*}\zeta^{-1}{\bf d})+
\frac 12{\bf d}\eta^{*}\zeta^{-1}{\bf d}-|{\bf d}
|^2\right].
\end{equation}
Then the photon distribution function for polymode light
in the squeezed Schr\"odinger cat state (even and odd)
 is given by the formula \cite{Nik}
\begin{equation}
{\cal P}_{n}^{\pm }(\beta )=
N_{\pm }^{2}(\beta )\left[|T_{n}(\beta )|^{2}+
|T_{n}(-\beta )|^{2}{\pm }\left(T_{n}^{*}(\beta )T_{n}(-\beta)+
T_{n}(\beta )T_{n}^{*}(-\beta )\right)\right].
\end{equation}
If the shift parameter $d=0$ the formula is simplified
\begin{equation}
{\cal P}_{2k}^{+}(\beta )=4N_{+}^{2}{\cal P}_{n}(\beta ),
\end{equation}
if we have equality
\begin{equation}
\sum _{i=1}^{N}n_{i}=2k,
\end{equation}
and for even states
\begin{equation}
{\cal P}_{2k+1}^{+}(\beta )=0,
\end{equation}
if one has
\begin{equation}
\sum _{i=1}^{N}n_{i}=2k+1,
\end{equation}
where ${\cal P}_{n}(\beta )$ is given by
the formula (3).For the light in the odd squeezed
Schr\"odinger cat state the photon distribution is
\begin{equation}
{\cal P}_{2k+1}^{-}(\beta )=4N_{-}^{2}{\cal P}_{n}(\beta ),
\end{equation}
if the indices satisfy the equality (26) and
\begin{equation}
{\cal P}_{2k}^{-}(\beta )=0,
\end{equation}
if the indices satisfy relation (24).Thus the squeezed Schr\"odinger
cat states if the shift parameter is equal to zero have highly
oscillating distribution function.The influence of shift
parameter decreases the oscillations of the distribution function.
For Hermite polynomials the following sum rule may be found
\cite{Ola2}
\begin{eqnarray}
&&\sum_{n_1=0}^{\infty}\ldots\sum_{n_N=0}^{\infty}\frac {\lambda_
1^{n_1}}{n_1!}\frac {\lambda_2^{n_2}}{n_2!}\ldots\frac {\lambda_N^{
n_N}}{n_N!}H_{n_1n_2\ldots n_Nn_1n_2\ldots n_N}^{\{{\bf R}\}}({\bf R}^{
-1}{\bf z})\nonumber\\
&=&\left[\det\left(\Lambda\Sigma_x{\bf R}+{\bf I}_{2N}\right)\right
]^{-\frac 12}\exp\left[\frac 12{\bf z}\left(\Lambda\Sigma_x{\bf R}
+{\bf I}_{2N}\right)^{-1}\Sigma_x\Lambda {\bf z}\right].\end{eqnarray}
\noindent Here ${\bf z}=(z_1,z_2,...z_{2N})$, the $2N$x$2N$ matrix $
\Sigma_x$ is the
$2N$-dimensional analog of the Pauli matrix $\sigma_x$,
and the diagonal $2N$x$2N$ - matrix $\Lambda $ has the matrix elements
$\lambda _{j}$ in j-th and (N+j)-th rows.
Let us consider the one mode case.Then the formula for the photon
distribution function in terms of Hermite polynomials of two variables
may be expressed in terms of usual Hermite polynomials

\[{\cal P}_n={\cal P}_0\frac {(T^2-4d)^{\frac n2}}{(2T+4d+1)^n}\sum_{
k=o}^n\left(\frac {4d-1}{\sqrt {T^2-4d}}\right)^k\frac {n!}{[(n-k
)!]^2k!}\]
\begin{equation}
\times\left|H_{n-k}\left(\frac {(T+1)z+\left[\sigma_{
pp}-\sigma_{qq}-2i\sigma_{pq}\right]z^{\star}}{\left\{(2T+4d+1)\left
[\sigma_{pp}-\sigma_{qq}-2i\sigma_{pq}\right]\right\}^{\frac 12}}\right
)\right|^2.
\end{equation}
Here the parameters $\sigma _{pp},\sigma _{qq},\sigma _{pq}$ are
matrix elements of the quadrature dispersion matrix $M$, $d$ is
determinant of this matrix and $T$ is the trace of the matrix.
The complex number $z$ is determined by the relation
\begin{equation}
z=\frac {1}{\sqrt 2}(<q>+i<p>).
\end{equation}
Formula (30) can be used also to illustrate the generalized
uncertainty relation (for the Gaussian states).  Indeed, it is
obvious that the probability to find $n$ photons must be nonnegative.
On the other hand, all but one terms in the right-hand side of eq.
(30) are positive independently on the concrete values of the
parameters determining the quantum state.  The only exception is
the term

\[\left(\frac {4d-1}{\sqrt {T^2-4d}}\right)^k.\]

\noindent Consequently, to guarantee the positiveness of
the photon distribution function for all
conceivable combinations of the parameters one should impose the
restriction $d\ge\frac 14$.This inequality is the Schr\"odinger
uncertainty relation (see, \cite{Sud},\cite{Dod89} ).

\section{Acknowledgments}
The author would like to acknowledge the University of
Maryland at Baltimore County for kind hospitality and
I.N.F.N.,Sez. di Napoli for support.

\end{document}